\documentstyle[prb,aps,twocolumn]{revtex} 
\begin{document}

\title{Superparamagnetic-like ac susceptibility behavior in a "partially disordered 
antiferromagnetic"  compound, Ca$_3$CoRhO$_6$}

\author{E.V. Sampathkumaran$^*$ and Asad Niazi}

\address{Tata Institute of Fundamaental Research, Homi Bhabha Road, 
Mumbai - 400 005, INDIA.}

\maketitle

\begin{abstract} 

We report the results of  dc and ac magnetization measurements as a function
of temperature (1.8 - 300 K) for the spin chain compound,  
Ca$_3$CoRhO$_6$, which has been recently reported to exhibit 
a partially disordered antiferromagnetic (PDAF) structure in the range 
30 - 90 K and spin-glass freezing below 30 K.  
We observe an unexpectedly 
large frequency dependence of ac susceptibility in the 
T range 30 - 90 K, typical of   
superparamagnets. In addition, we find that there is no difference 
in the isothermal remanent magnetization behavior for the two regimes below 90 K. 
These findings call for  more investigations to understand the magnetism of this compound. 

\end{abstract}
\vskip1cm
{PACS  numbers: 75.50.-y; 75.30.Cr; 75.50.Lk}
\vskip0.5cm
$^*$E-mail address: sampath@tifr.res.in
\vskip1cm

\maketitle

Recently, the spin-chain compound, Ca$_3$CoRhO$_6$ (Ref. 1), 
has started getting attention due to the fact this compound exhibits 
a novel type\cite{2,3} of magnetic phase transition. This compound 
crystallizes in a K$_4$CdCl$_6$ type rhombohedral structure (space group R$\bar3$c). With 
this structure, in this compound, there are one-dimensional chains of alternating 
face-sharing CoO$_6$ trigonal prisms and RhO$_6$ octahedra. Ca cations
separate these chains, and the magnetic ions form a triangular lattice
with an interchain spacing of 5.313 $\AA$. The dc magnetic susceptibility ($\chi$) 
as well as neutron diffraction data reveal\cite{1,2} that there are two 
magnetic transitions, one at 90 K (T$_1$) and the 
other at 30 K (T$_2$). In the magnetically ordered state, the magnetic 
ions along the chain 
couple ferromagnetically, whereas the interchain nearest neighbour interaction is 
antiferromagnetic. This type of interaction may cause magnetic frustration,
as this compound is intrinsically a triangular lattice. As a consequence
of this geometrical frustration, between T$_1$ and T$_2$, 2/3 of the ferromagnetic chains (present at the
corners of the hexagon) have been reported to 
couple antiferromagnetically with each other and the other 1/3 (at the centre
of the hexagon) are proposed to be left
 incoherent with the other chains (see Fig. 1a of Ref. 3). 
There are only  few compounds known 
with this
kind of triangular lattice of antiferromagnetic spin chains, e.g., CsCoCl$_3$
and CsCoBr$_3$\cite{4,5}. However, the magnetism of this compound is proposed\cite{3} to be
unusual in the sense that, below T$_2$, the incoherent chains appear to 
freeze in a disordered
state, whereas in the Cs compounds, a ferrimagnetic state is attained below the lower 
transition temperature. However, the ferrimagnetic phase can be 
obtained even in the present material by an application of a magnetic field (H) of the
order of 20 to 30 kOe. In other words, this compound is characterized by an 
interesting magnetic phase diagram\cite{2}. 

While we believe that this novel magnetic compound will attract future attention for various
investigations, it
is of interest to look for further characteristics of such a complex 
magnetic material, in particular, to compare the anomalies due to the disordered
nature of the magnetism in the two temperature ranges below 90 K. With 
this motivation, we have carried out magnetic measurements on this material, the
results of which are reported in this article. The results apparently raise interesting questions. 

The compound Ca$_3$CoRhO$_6$ in the polycrystalline form was prepared by 
solid state route. Stoichiometric amounts of high purity ($>$99.9\%) CaCO$_3$, 
CoO and Rh powder were thoroughly mixed. Then the mixture was 
calcined at 900 C for one day. Subsequently, the preacted powder was
finely ground, pelletized and  heated at 1200 C for about 10 days with few 
intermediate grindings. The x-ray diffraction pattern confirmed that the 
sample is a single phase (a= 9.202~$\AA$ ~ and ~ c= 10.730~$\AA$). The dc $\chi$ measurements were performed 
in the range 1.8 - 300 K at different fields (1, 30 and 50 kOe) both for 
zero-field-cooled (ZFC) as well as field-cooled (FC) state of the specimens 
employing a commercial (Quantum Design) superconducting 
quantum interference device (SQUID). The same magnetometer was employed to take ac $\chi$ data
(1.8 - 300 K) at various frequencies in zero field as well as in the presence of two dc 
fields (5 and 30 kOe). Isothermal  magnetization (M) and remanent magnetization (M$_{IRM}$) behavior  
were also tracked at 5 and 62 K employing a commercial magnetometers both by  
vibrating sample magnetometer (VSM, Oxford Instruments) and SQUID and the results obtained
from both the instruments agree quite well.        

The T-dependence of dc $\chi$ measurements are shown in Fig. 1. There is a sharp drop  of
$\chi$ above about 35 K for ZFC-curves for all H  as though there is a magnetic 
transition at this temperature. ZFC $\chi$-T plots tend to merge below about 20 K 
for all the values of H. However, FC-curves at low temperatures 
show a different behavior in the sense that the values are found to be H-dependent,
though the plots tend to a constant value below 20 K in all cases. Above about 35 K, $\chi$ values are 
field-dependent for both the ZFC and FC measurements and the features are suggestive of another transition in the range 90 - 100 K. 
 As noted earlier\cite{1}, the plot of inverse
$\chi$ versus T (Fig. 1, inset) is highly non-linear in the paramagnetic state (above 90 K). 
The Curie-Weiss temperature inferred from the near-linear range of 225 - 300 K
is found to be about 175 K, the positive sign indicating strong ferromagnetic interaction; this
value however is marginally higher than that reported (150 K) in Ref. 1.

The findings are otherwise in good agreement with those reported in Ref. 1. There is a distinct difference\cite{1} between the plots of 
isothermal M at 5 and 62 K (Fig. 1, inset) in the sense that  there is a plateau in the H range 
20 - 40 kOe at 62 K
due to ferrimagnetic alignment of the spins. There is a significant hysteresis
in the plot for 5 K resulting in a remanence of M ({\it vide infra}) after reducing H to zero, which is in 
agreement with the proposed spin-glass behavior. We have observed similar magnitudes of remanence even at 62 K, which is not obvious from Fig. 1 inset, due to the compression of the scale on the y-axis. The ZFC-FC $\chi$ curves obtained in the presence of 
 1 kOe field tend to bifurcate at a temperature close to 70 K; however, applications of higher fields lower the temperature at which this bifurcation occurs, implying thereby that at much smaller fields (in few gausses) this feature may occur at temperatures higher than 70 K. There is 
a tendency for a broad peak  in temperature range 100 to 150 K in $\chi$ versus T plots for H= 1 and 30 kOe, but it is cut off around 100 K due to the onset of magnetic ordering. It is to be noted this peak is clearly visible in the data for H= 50 kOe. This maximum in $\chi$ resembles that of Bonner and Fischer's prediction for one-dimensional magnetism\cite{6}. Alternatively, short-range correlations also may be responsible for this feature.

We now present the results of ac $\chi$ measurements (see Fig. 2). It is clearly seen that, in the zero-field data, 
there is a well-defined peak at 50 K in the real part ($\chi$$\prime$) of ac $\chi$ typical of 
spin-glasses. However, what is remarkable is that the anomalies associated with magnetic ordering
start building at temperatures as high as 90 K. Therefore, it is very difficult to associate the ac $\chi$
anomalies (see further below) to the transition at T$_2$. This observation may be correlated to the  ZFC-FC divergence at a temperature close to 90 K  at  low fields (say, at 1 kOe) as discussed above.  The fact 
that the $\chi$$\prime$ peak arises from some kind of spin-glass freezing appears to  be endorsed by the
observation of an upturn in the imaginary part ($\chi$$\prime$$\prime$) of the ac $\chi$ as well (prominent below about 75 K). 
One does not expect such a feature at the magnetic transition temperature due to long range 
magnetic ordering\cite{7}. It is also remarkable that the width of the $\chi$$\prime$-feature is so large that it spans entire temperature range between T$_1$ and T$_2$. What is puzzling is the large frequency dependence of the temperature at which the  peak appears, for instance, from 50 K at 1 Hz to 70 K at 1 kHz in the
$\chi$$\prime$ data, in sharp contrast to that known in canonical spin glasses. Quantitatively speaking, 
the magnitude of the factor\cite{7}, $\Delta$T$_f$/T$_f$$\Delta$(log$\omega$), turns out to be as large as 
about 0.1, which is typically seen in superparamagnetic materials\cite{7}. 
The peak temperature in the $\chi$$\prime$ vs T plot 
typically represents the spin-freezing temperature (T$_f$). The value of this factor in canonical spin glasses is known to be below 0.01. It is to be noted
that the values of $\chi$ are also practically frequency-independent below 30 K. We have also taken
the data in the presence of a H of 5 and 30 kOe. While the ac chi features for 5 kOe resemble those of zero-field data, 
for H= 30 kOe data, the $\chi$$\prime$ cusp vanishes and the peak gets 
broadened by the field with a significantly reduced intensity; also the $\chi$$\prime$$\prime$-anomaly is completely depressed. This finding is consistent with 
the proposed phase diagram\cite{2} in the sense that the disordered magnetic state is destroyed 
at  a field of 30 kOe.

We have also probed  the magnetic relaxation behavior in the two temperature ranges, by measuring
at 5 and 62 K (Fig. 3). For this purpose, we have zero-field-cooled
the specimen to respective temperatures, switched on the field of 5 kOe, waited for 5 mins and then
the decay of M$_{IRM}$ was tracked as a function of time (t) for about an hour after the field was switched off. We observed that M$_{IRM}$ dropped to a value below the detection level of the instrument in the paramagnetic state (say, at 120 K) as expected. However, the values at 5 and 62 K are significant (in agreement with the M versus H behavior discussed above) and found to undergo slow decay with t. A quantitative analysis of the data to an exponential form popularly noted for canonical spin glasses is found to hold good for a narrow range of the plot (below about 1000 sec), but deviations occur 
for M$_{IRM}$ curve at higher t values. It appears that, barring low-t region, M$_{IRM}$ seems to vary logarithmically as shown in Fig. 3 with comparable values of the coefficient of the logarithmic term.  It thus appears that a superimposition of exponential (for low t values) and logarithmic (at higher t values) terms describe the relaxation process. As the actual form of relaxation is  a matter of great discussion even in
conventional spin glasses,\cite{7} a further discussion on this aspect for the present complex material is not desirable. What is important to note is that, while the slow relaxation of  M$_{IRM}$ at 5 K is consistent
with the spin-glass freezing proposed in Ref. 3, similar behavior at 62 K is unexpected on the basis of the analysis of the neutron diffraction data. If the interpretation of previous results\cite{3} are correct, one can argue that coexisting antiferromagnetic chains and paramagnetic chains can also give rise to this behavior which by itself will be an interesting conclusion. Alternatively, on the basis of the similarities of the relaxation behavior at 5 and 62 K quantitatively evidenced by a comparable values of the coefficients of the logarithmic term (see Fig. 3), one is also
tempted to propose that the  dynamics  in the entire temperature range below 90 K observed here is the one due to superparamagnetic clusters formed between T$_1$ and T$_2$, the dynamics of which apparently gets frozen below T$_2$.

To conclude, we have observed a large frequency dependence of 
ac susceptibility in Ca$_3$CoRhO$_6$ in a temperature range in which 
a partially disordered antiferromagnetic phase has been previously proposed. If the interpretation of the neutron diffraction data\cite{2} for the range 30-90 K is correct, the present results raise a very interesting question how the coexistence of antiferromagnetic and incoherent magnetic chains result in 
superparamagnetic-like behavior. It is also surprising that the magnetic relaxation behavior in the two temperature regimes in the magnetically ordered state is found to be similar, which prompts us to propose that the dynamics of the superparamagnetic clusters formed in the range 30-90 K freeze below about 30 K.  We hope that this work motivates  further microscopic investigations for a better understanding of the magnetism of this compound.

The help of Kartik K Iyer during measurements
is acknowledged. 


\vskip2cm
\begin{figure}
\vskip5mm
\caption{Dc magnetic susceptibility ($\chi$) as a function of temperature for 
Ca$_3$CoRhO$_6$ measured at different fields. The transition temperatures, T$_1$ and T$_2$, are marked. 
Insets show (a) inverse $\chi$ 
(H= 50 kOe) as a function of T to highlight paramagnetic behavior with the lines through the data points serving as a 
guide to the eyes and (b) isothermal magnetization at 5 and 62 K. For T= 5 K, the lower curve corresponds to increasing H and the upper one to decreasing H. }
\end{figure} 
\vskip1cm

\vskip2cm
\begin{figure}
\vskip5mm
\caption{Real ($\chi$$\prime$) and imaginery ($\chi$$\prime$$\prime$) parts of ac susceptibility
for Ca$_3$CoRhO$_6$, measured at different frequencies with an ac driving field
of 1 Oe. The  ac $\chi$ behavior in the presence of a magnetic field of 5 kOe at two frequencies are  
also shown. The data for the presence of 30 kOe at various frequencies nearly overlap and hence shown only for
one frequency (with the data for $\chi$$\prime$$\prime$ being featureless not shown in the plot). The curves shown are obtained by drawing lines through the experimental data points.}
\end{figure}
\vskip1cm

\vskip2cm
\begin{figure}
\vskip5mm
\caption{Isothermal remanent magnetization behavior at 5 and 62 K for Ca$_3$CoRhO$_6$, 
obtained as described in the text. The continuous lines represent least squares fit to a logarithmic function mentioned in the figure.}
\end{figure}
\vskip1cm

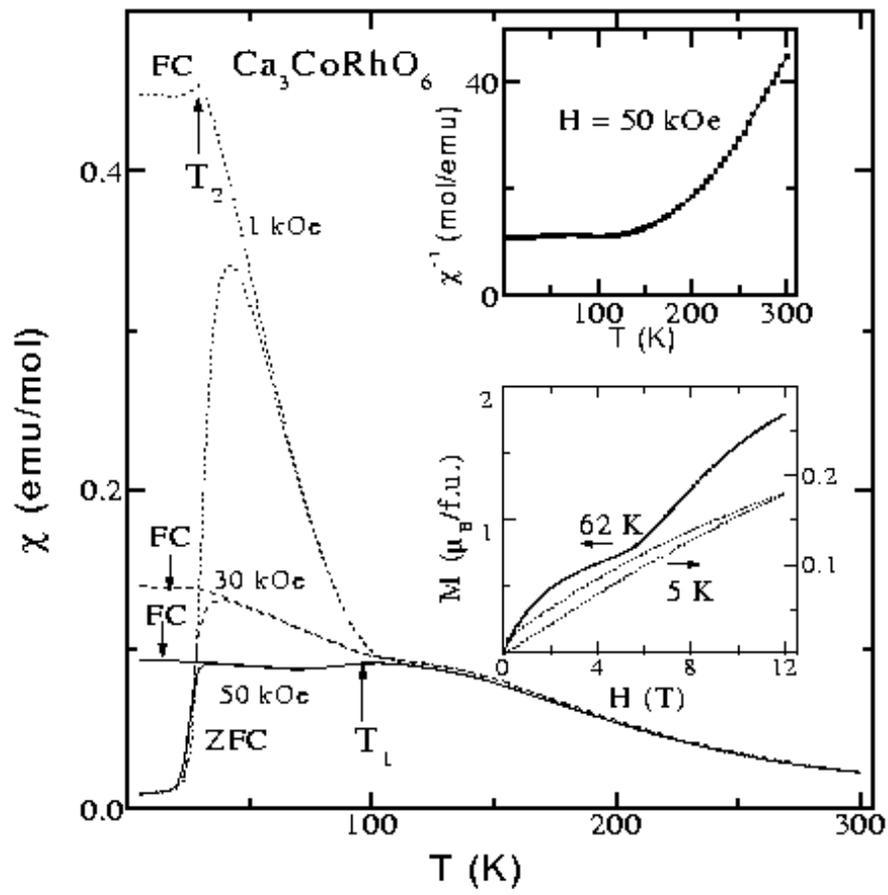
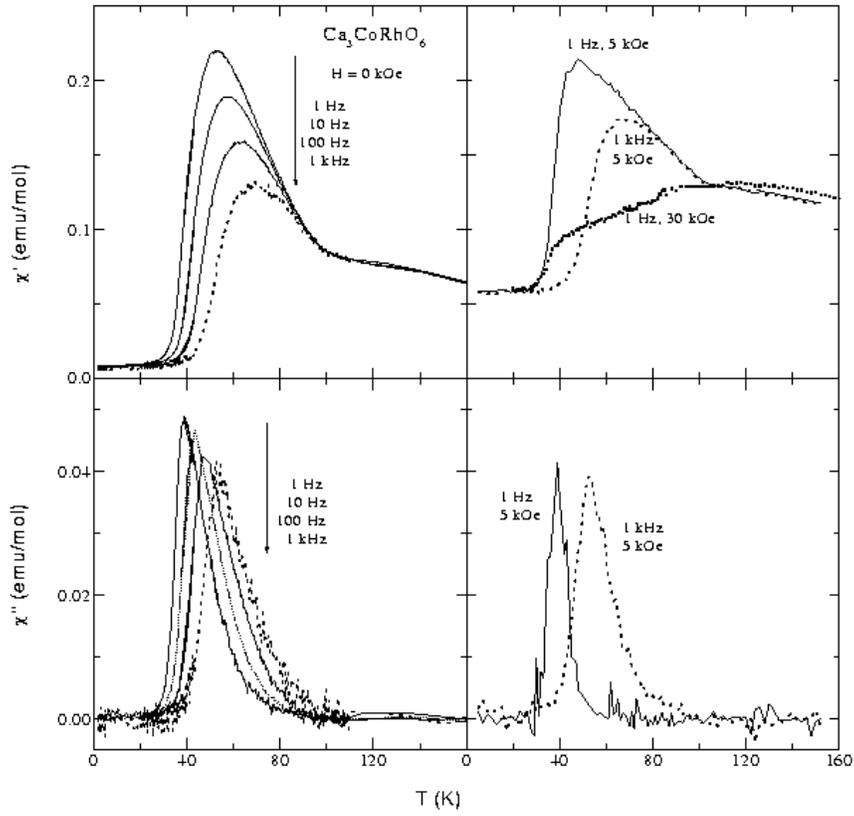
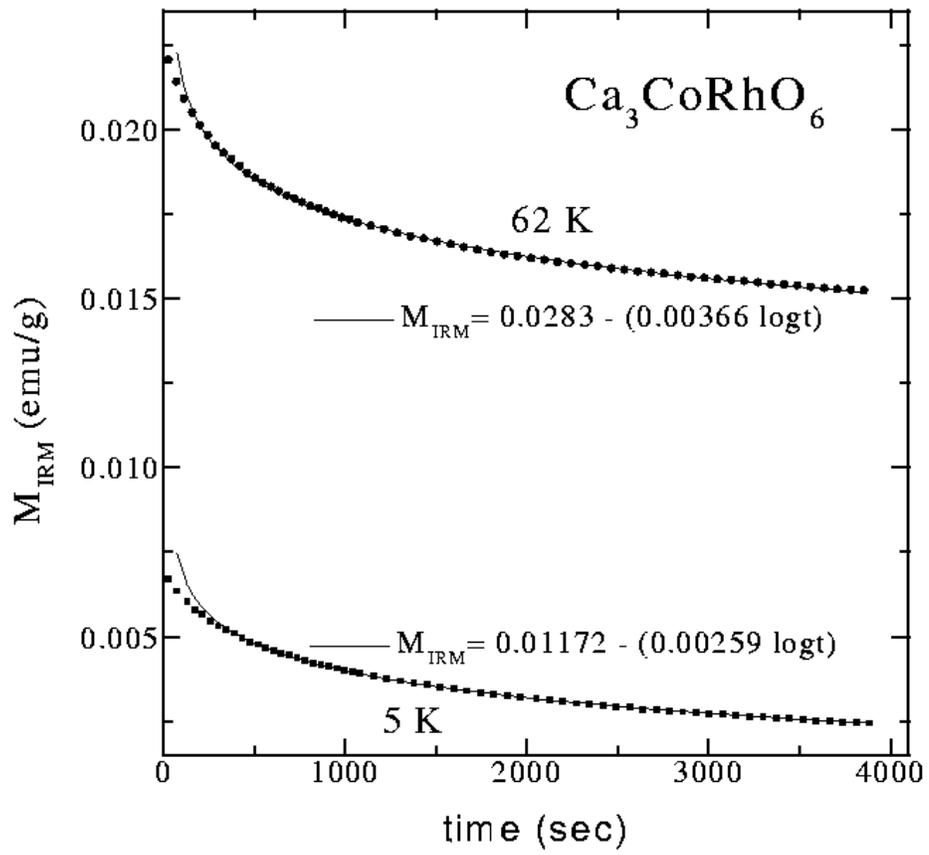
\end{document}